\documentclass[12pt]{article}
\usepackage{graphicx}
\usepackage{lineno}

\newcommand\pubnumber{ }
\newcommand\pubdate{ }

\def\mcgill{$^a$Department of Physics, McGill University, Rutherford Physics Building, 3600 rue University, Montr\'eal, Qu\'ebec, H3A~2T8, CANADA}
\def\bnl{$^b$Physics Department, Brookhaven National Lab, Building 510A, Upton, NY, 11973, USA}
\def\santiago{$^c$Departamento de F\'isica de Part\'iculas and IGFAE, Universidade de Santiago de Compostela, E-15706 Santiago de Compostela, Galicia-Spain}
\def\sanpaulo{$^d$Instituto de F\'isica - Universidade de S\~ao Paulo, Rua do Mat\~ao Travessa R, no. 187, 05508-090, Cidade Universit\'aria, S\~ao Paulo, Brasil}

\def\Title#1{\begin{center} {\Large #1 } \end{center}}
\def\Author#1{\begin{center}{ \sc #1} \end{center}}
\def\Address#1{\begin{center}{ \it #1} \end{center}}

\newcommand\pubblock{\rightline{\begin{tabular}{l} \pubnumber\\
         \pubdate  \end{tabular}}}
\newenvironment{Abstract}{\begin{quotation}  }{\end{quotation}}
\newenvironment{Presented}{\begin{quotation} \begin{center} 
             PRESENTED AT\end{center}\bigskip 
      \begin{center}\begin{large}}{\end{large}\end{center} \end{quotation}}
\def\Acknowledgements{\bigskip  \bigskip \begin{center} \begin{large}
             \bf ACKNOWLEDGEMENTS \end{large}\end{center}}

\def\beq{\begin{equation}}
\def\eeq#1{\label{#1}\end{equation}}
\def\eeqn{\end{equation}}

\def\beqa{\begin{eqnarray}}
\def\eeqa#1{\label{#1}\end{eqnarray}}
\def\eeqan{\end{eqnarray}}

\let\bar=\overbar

\def\Dslash{\not{\hbox{\kern-4pt $D$}}}
\def\dslash{\not{\hbox{\kern-2pt $\del$}}}

\def\msb{{\bar{\ssstyle M \kern -1pt S}}}

\begin{document}
\begin{titlepage}
\pubblock

\vfill
\Title{Dilepton emission in high-energy heavy-ion collisions with dissipative hydrodynamics}
\vfill
\Author{Gojko Vujanovic$^{a}$, Gabriel S. Denicol$^{a,b}$, Chun Shen$^{a}$, Matthew Luzum$^{c,d}$, Bj\"orn Schenke$^{b}$, Sangyoung Jeon$^{a}$, and Charles Gale$^{a}$}
\Address{\mcgill}
\Address{\bnl}
\Address{\santiago}
\Address{\sanpaulo}
\vfill
\begin{Abstract}
In this contribution we study the effects of three transport coefficients of dissipative hydrodynamics on thermal dilepton anisotropic flow observables. The first two transport coefficients investigated influence the overall size and growth rate of shear viscous pressure, while the last transport coefficient governs the magnitude of net baryon number diffusion in relativistic dissipative fluid dynamics. All calculations are done using state-of-the-art 3+1D hydrodynamical simulations. We show that thermal dileptons are sensitive probes of the transport coefficients of dissipative hydrodynamics.   
\end{Abstract}
\vfill
\begin{Presented}
Twelfth Conference on the Intersections of Particle and Nuclear Physics (CIPANP 2015)\\
 $\phantom{}$ \\
Vail, Colorado, USA,  May 19--24, 2015
\end{Presented}
\vfill
\end{titlepage}
\def\thefootnote{\fnsymbol{footnote}}
\setcounter{footnote}{0}
%

\section{Introduction}
A hydrodynamical model of the medium created in relativistic heavy-ion collisions allows to constrain various transport properties of QCD. At high collisions energies such as those at the Relativistic Heavy Ion Collider (RHIC) at Brookhaven National Laboratory (BNL) or at the Large Hadron Collider (LHC) at CERN, focus is given to extracting viscous transport coefficients, such as shear and bulk viscosity, of nuclear media. At lower collisions energies | such as those probed by the Beam Energy Scan (BES) Program at RHIC or at the upcoming the Compressed Baryon Matter (CBM) Experiment at the Facility for Antiproton and Ion Research (FAIR) -- hydrodynamical simulations become more sensitive to conserved charges their diffusion; the prime example being net baryon number. Hence, at the highest collisions energy produced at RHIC $\sqrt{s_{NN}}=200$ GeV, we will explore the effects of two transport coefficients governing the size and growth rate of the shear viscous pressure, shear viscosity and shear relaxation time, via their influence on the anisotropic flow of thermal dileptons. Similarly, at the lowest collision energy probed by the BES $\sqrt{s_{NN}}=7.7$ GeV, we will investigate the effects of net baryon number density and its diffusion on dilepton emission. 

The interest in describing lower beam energy nucleus-nucleus collisions via hydrodynamics was invigorated in recent years, owing to developments in Lattice QCD ($\ell$QCD), which can describe the Equation of State (EoS) of nuclear media at high temperature (T) and finite net baryon chemical potential $\mu_B$. The description of the $\ell$QCD EoS at finite $\mu_B$ currently relies on using the Taylor expansion of the grand canonical partition function around $\mu_B=0$ \cite{Borsanyi:2011sw, Bazavov:2012jq}.\footnote{However, a Taylor expansion around $\mu_B=0$ cannot capture the physics of phase transitions that may occur at a certain point in the $T$--$\mu_B$ plane, as the EoS would become a non-analytic function around a phase transition. Hence other method, e.g. complex Langevin dynamics \cite{Sexty:2014zya}, bear promise to extract the EoS at high T or $\mu_B$.} So, these proceedings will use two equations of state: at high energies, the canonical formalism is used \cite{Huovinen:2009yb}, while at lower beam energies the grand canonical formalism is employed instead \cite{Aki_priv}.

Let it be made clear at this point that given how recent the equation of state at finite $\mu_B$ is, and the general lack studies regarding the size various transport coefficients of dissipative hydrodynamics at lower beam energies (especially their $\mu_B$-dependence), we will be focusing on trends observed in the dilepton results as the $\mu_B$-dependent EoS and baryon diffusion are dialed in the hydrodynamical evolutions.    

\section{Hydrodynamical equations}
Throughout these proceedings, the second order Israel-Stewart hydrodynamical equation of motion being solved are:
\begin{eqnarray}
\partial_\mu T^{\mu\nu} &=& \partial_\mu \left[T^{\mu\nu}_{(0)} + \pi^{\mu\nu}\right] = \partial_\mu \left[\varepsilon u^\mu u^\nu -\Delta^{\mu\nu} P + \pi^{\mu\nu}\right]\\
\partial_\mu J^{\mu}_{B} &=& n_B u^\mu + V^\mu 
\end{eqnarray}
where $T^{\mu\nu}$ is the stress-energy tensor, $u^\mu$ is the fluid 4-velocity, $P$ is the pressure and $\Delta^{\mu\nu}=g^{\mu\nu}-u^\mu u^\nu$, $g^{\mu\nu}$ is the Minkowski metric with signature $(+,-,-,-)$, $\pi^{\mu\nu}$ is the shear viscous pressure tensor (or simply the shear-stress tensor), $n_B$ is the net baryon number density, and $V^\mu$ is the net baryon number diffusion 4-vector (or simply the baryon diffusion). The shear-stress tensor and the baryon diffusion further satisfy the following relaxation equations:
\begin{eqnarray}
\tau_\pi \Delta^\mu_\alpha \Delta^\nu_\beta u^\sigma \partial_\sigma \pi^{\alpha \beta} + \pi^{\mu\nu} &=& 2\eta\sigma^{\mu \nu} - \frac{4}{3} \tau_\pi \pi^{\mu \nu} \theta \label{eq:pi_munu_evol}\\
\tau_V \Delta^\mu_\nu u^\sigma \partial_\sigma V^\nu + V^\mu &=& \kappa \nabla^\mu \alpha_0- \tau_{V} V^\mu \theta - \frac{3}{5} \tau_V \sigma^{\mu\nu} V_{\nu}
\end{eqnarray}   
where $\tau_\pi$ and $\tau_V$ are the relaxation times for the shear-stress tensor and the baryon diffusion, respectively, while $\sigma^{\mu \nu}=\Delta^{\mu\nu}_{\alpha\beta}\partial^\alpha u^\beta$ with $\Delta^{\mu\nu}_{\alpha\beta}=\frac{1}{2}\left[\Delta^\mu_\alpha \Delta^\nu_\beta + \Delta^\mu_\beta \Delta^\nu_\alpha \right]-\frac{1}{3}\Delta^{\mu\nu}\Delta_{\alpha\beta}$, $\theta=\partial_\mu u^\mu$, and $\alpha_0=\frac{\mu_B}{T}$. Finally, $\eta$ is the shear viscosity and $\kappa$ is the net baryon number conductivity. Terms containing $\frac{4}{3} \tau_\pi$, and $\frac{3}{5} \tau_V$, along with the term $\tau_V$ in front of $V^\mu\theta$, can be derived from solving the Boltzmann equation in the 14-moment approximation. 

At high energies $\sqrt{s_{NN}}=200$ GeV we will set $n_B$ to zero. In that case, we will study the effects of $\tau_\pi$ and $\eta$ on the evolution of the medium. Their effects will be investigated by varying them as follows: 
\begin{eqnarray}
\tau_\pi &=& b_\pi \frac{\eta}{\varepsilon+P}\label{eq:tau_pi}\\
\frac{\eta}{s} &=& m\left(\frac{T}{T_{tr}}-1\right)+\frac{1}{4\pi}\label{eq:eta_s}
\end{eqnarray}
where 3 values of $b_\pi$ and $m$ are chosen, namely $b_\pi=5,10,20$ and $m=0,0.2427,0.5516$, while $T_{tr}=0.18$ GeV. More precisely, while the effects of $\tau_\pi$ are being studied, $m=0$ is being kept throughout, while when the effects of $\frac{\eta}{s}$ are being studied $b_\pi=5$. 

At a beam energy of $\sqrt{s_{NN}}=7.7$ GeV, the goal is to explore the influence of $n_B$ and $V^\mu$ on the hydrodynamical evolution. Indeed, we will only be interested in trends these two degrees of freedom induce on the evolution of the medium, with more detailed studied being reserved for a future publication. For that purpose, we choose $\tau_V=\frac{0.2}{T}$, $\kappa=0.2\frac{n_B}{\mu_B}$, which are inspired from the AdS/CFT calculation in Ref. \cite{Natsuume:2007ty}. 

\section{Initial and freeze-out conditions}
Given the two sets of collision energies being explored, i.e. $\sqrt{s_{NN}}=200$ GeV and $\sqrt{s_{NN}}=7.7$ GeV, two sets initial conditions will be discussed. For high energy collisions we use the Monte Carlo (MC) Glauber model, while at lower beam energy an average of 1000 MC Glauber initial distributions is used. In both cases however, no fluctuation are present in the longitudinal direction (i.e. along the beam direction) hence the initial energy density profile can be decomposed into a longitudinal and a transverse piece. To do so, we introduce hyperbolic coordinates $(t,z)=(\tau\cosh\eta_s,\tau\sinh\eta_s)$, where $\tau$ is the proper time $\tau=\sqrt{t^2-z^2}$, and $\eta_s$ is the space-time rapidity $\eta_s=(1/2)\ln\left[(t+z)/(t-z)\right]$. Thus, the energy density profile $\varepsilon \left( \tau _{0},x,y,\eta \right) = g(\eta_s) \varepsilon _{T}\left( x,y\right)$ where, at $\sqrt{s_{NN}}=200$ GeV, we take a $g(\eta_s)$ profile that is approximately boost invariant near mid-rapidity, falling off like a Gaussian at large rapidities:
\begin{eqnarray}
g(\eta_s)=\exp \left[ -\frac{\left(
\left\vert \eta \right\vert -\eta _{\mathrm{flat}}/2\right) ^{2}}{2\eta
_{\sigma }^{2}}\theta \left( \left\vert \eta \right\vert -\eta _{\mathrm{flat%
}}/2\right) \right]
\end{eqnarray}
with $(\eta _{\mathrm{flat}},\eta _{\sigma })=(5.9,0.4)$, thus obtaining a good description of the $\frac{dN}{dy}$ distribution of charged hadrons at 20-40\% centrality. At $\sqrt{s_{NN}}=7.7$ GeV,
\begin{eqnarray}
g(\eta_s) &=& N \left\{ \Theta(|\eta_s|-\eta_{s,0}) \exp\left[-\frac{(|\eta_s|-\eta_{s,0})^2}{2\Delta\eta^2_{s,1}}\right]\right.\nonumber\\
    &+& \left. \left[1-\Theta(|\eta_s|-\eta_{s,0})\right]\left[A+\left(1-A\right) \exp\left[-\frac{(|\eta_s|-\eta_{s,0})^2}{2\Delta\eta^2_{s,2}}\right] \right] \right\},\\
N   &=& \frac{1}{\sqrt{2\pi}\Delta\eta_{s,1}+(1-A)\sqrt{2\pi}\Delta\eta_{s,2} + 2A\eta_{s,0}} 
\end{eqnarray}
where $g(\eta_s)$ is used to initialize both the energy density and the net baryon density ($n_B$) in the longitudinal direction, while $N$ is chosen such that $\int d\eta_s g(\eta_s) = 1$. The parameters $\eta_{s,0}=2.09$, $A=0.8$, $\Delta\eta_{s,1}=0.7$, $\Delta\eta_{s,2}=1$ were chosen so that the net proton rapidity distribution $\frac{dN}{dy}$ is reproduced at $\sqrt{s_{NN}}=7.7$ GeV and 0-80\% centrality. The transverse direction is further decomposed into a contribution from binary nucleon-nucleon collisions and wounded nucleons, the latter being computed according to the Glauber model \cite{Miller:2007ri}. 

We always take the initial flow profile to be zero, i.e., $u^{\mu}(\tau_0)=(1,0,0,0)$ in $(\tau,\eta)$ coordinates, where $\tau_0=0.4$ fm/c at $\sqrt{s_{NN}}=200$ GeV, while $\tau_0=0.6$ fm/c at $\sqrt{s_{NN}}=7.7$ GeV. The initial shear-stress tensor and baryon diffusion vector are $\pi^{\mu\nu}(\tau_0)=0$ and $V^{\mu}(\tau_0)=0$. The energy density being initialized, we now turn our attention towards the initial net baryon number density $n_B$. Indeed, we initialize $n_B$ as:
\begin{eqnarray}
n_B(x,y,\eta_s) = F g(\eta_s) n^{\bot}_B(x,y) =  F g(\eta_s) 2 T_{\rm Au}(x,y)
\end{eqnarray}
where $T_{\rm Au}=\int dz \rho_{\rm Au} ({\bf x})$ is the nuclear thickness function of nucleus where $\rho_{\rm Au} ({\bf x})$ is the Woods-Saxon distribution of the gold nucleus. The overall normalization $F$ is tuned such that the number of participants $N_{\rm part}$ obtained from the Glauber model is reproduced. 

At this point all initial conditions have been specified. The hydrodynamical evolution is frozen-out a constant $T_{FO}=145$ MeV at $\sqrt{s_{NN}}=200$, while a constant energy density of 0.1 GeV/fm$^3$ is used at $\sqrt{s_{NN}}=7.7$ GeV. Indeed, we found that freezing-out at a constant energy density instead of temperature yields a better fit to hadrons at $\sqrt{s_{NN}}=7.7$ GeV. The next step is to present the dilepton rates being used throughout these proceedings, before discussing results. 

\section{Dilepton production rates}
There are two important sources of thermal dileptons: in the hadronic medium (HM) the dominant source of thermal dileptons comes from in-medium vector mesons, while in the partonic sector an important source of dileptons comes from the quark--antiquark annihilation in the Born approximation. There are more sophisticated calculations in the quark gluon plasma (QGP) sector, such as those in Refs. \cite{Ghisoiu:2014mha,Ghiglieri:2014kma}, however these two calculations are currently not amenable to a dissipative description of the medium and hence will not be considered at this time. 

In the hadronic sector the dilepton production proceeds through a direct decay of in-medium vector mesons into dileptons, giving the dilepton rate:
\begin{eqnarray}
\frac{d^4 R}{d^4 q} = - \left[1+\frac{2m^2_\ell}{q^2}\right]\left[1-\frac{4m^2_\ell}{q^2}\right]^{1/2} \frac{\alpha^2}{\pi^3} \frac{1}{q^2}  \sum_{V=\rho,\omega,\phi} \frac{m^4_V}{g_V^2} \frac{{\rm Im}D^R_V(q;\beta,\mu_B)}{\left(e^{\beta q^0}-1\right)}
\end{eqnarray} 
where $m_\ell$ is the mass of the lepton, $q^2=q^\mu q_\mu$ and $q^\mu$ is the 4-momentum of the virtual photon, $\alpha=1/137$ is the electromagnetic coupling, $m_V$ is the mass of the vector meson, $g_V$ is the strength of the coupling between the vector meson and the photon, $\beta=\frac{1}{T}$ is the inverse temperature, while ${\rm Im}D^R_V$ is the imaginary part of the vector meson propagator, which contains its in-medium mass and width properties. The computation of ${\rm Im}D^R_V$ is involved and can be found in Ref. \cite{Vujanovic:2013jpa}. 

The Born dilepton rate in the QGP is: 
\begin{eqnarray}
\frac{d^4 R^{\ell^+ \ell^-}}{d^4 q} &=& \frac{\alpha^2_{EM}}{6\pi^4} \left[1+\frac{2m^2_\ell}{q^2}\right]\left[1-\frac{4m^2_\ell}{q^2}\right]^{1/2} \frac{1}{e^{\beta q^0}-1}\times H(q;\beta,\mu_B)
\end{eqnarray}
where
\begin{eqnarray}
H (q;\beta,\mu_B) = \left\{ 1-\frac{1}{\beta \left|{\bf q}\right|} \ln\left[\frac{1+e^{-\frac{\beta\left(q^0-|{\bf q}|+\mu_B/3\right)}{2}}}{1+e^{-\frac{\beta\left(q^0+|{\bf q}|+\mu_B/3\right)}{2}}}\right] - \frac{1}{\beta \left|{\bf q}\right|} \ln\left[\frac{1+e^{-\frac{\beta\left(q^0-|{\bf q}|-\mu_B/3\right)}{2}}}{1+e^{-\frac{\beta\left(q^0+|{\bf q}|-\mu_B/3\right)}{2}}}\right]\right\}.\nonumber\\
\end{eqnarray}

The dissipative correction $\delta R$ to the dilepton rate accounting for baryon diffusion, in both the QGP and the HM sectors, is computed in a similar fashion to the shear viscous correction derived in Ref. \cite{Vujanovic:2013jpa}. That is, the dissipative correction $\delta n$ to the thermal distribution function takes the form: 
\begin{eqnarray}
n_{\rm total}(p\cdot u) = n(p\cdot u-b_i\mu_B) + C n(p\cdot u)(1 \pm n(p\cdot u)) \left[\frac{n_B T}{\varepsilon+P} - \frac{b_i T}{p\cdot u}\right] \frac{p^\mu V_\mu}{T\hat{\kappa}}\nonumber\\
\end{eqnarray} 
where $n$ is the thermal Bose-Einstein/Fermi-Dirac distribution with $p^\mu$ being the 4-momentum of the Boson/Fermion, $\hat{\kappa}=\frac{\kappa}{\tau_V}$, $C=1$ for simplicity, while 
\begin{eqnarray}
b_i=\left\{
\begin{array}{ll}
-1 & \mathrm{if\,} i\, \mathrm{is\, a\, antibaryon}\\
1 & \mathrm{if\,} i\, \mathrm{is\, a\, baryon}\\ 
0 & \mathrm{otherwise}
\end{array}\right.
\quad 
b_i=\left\{
\begin{array}{ll}
-\frac{1}{3} & \mathrm{if\,} i\, \mathrm{is\, a\, antiquark}\\
\frac{1}{3} & \mathrm{if\,} i\, \mathrm{is\, a\, quark}\\ 
0 & \mathrm{otherwise.}
\end{array}\right.
\end{eqnarray} 
%

\section{Effects of $\tau_\pi$ and $\frac{\eta}{s}(T)$}\label{sec:res_tau_pi}
\begin{figure}[!h]
\begin{center}
\includegraphics[width=0.496025\textwidth]{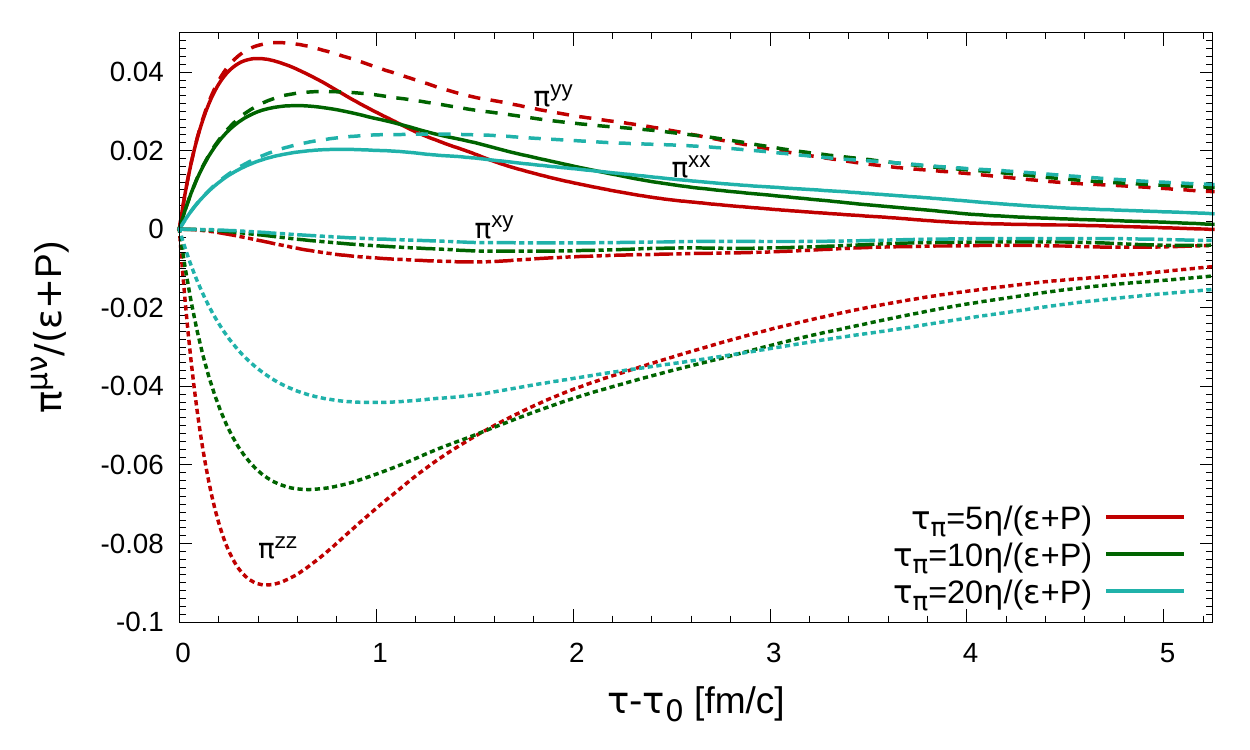}
\includegraphics[width=0.496025\textwidth]{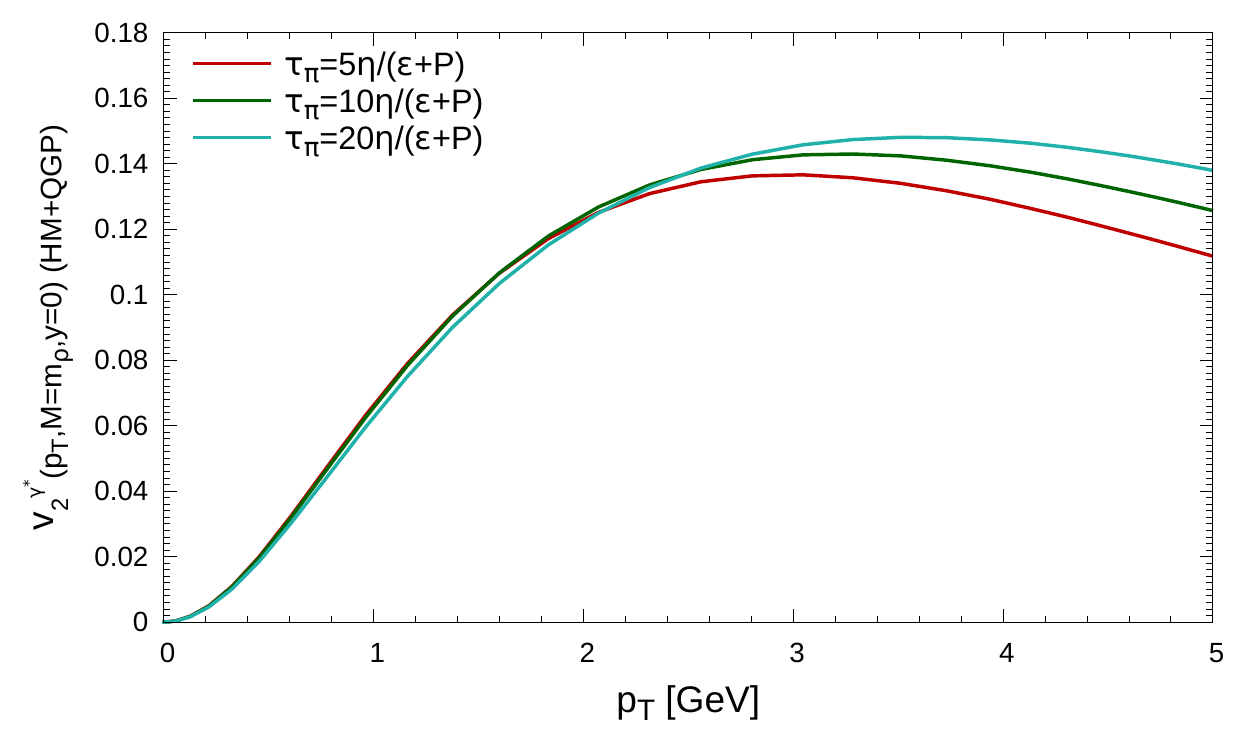}
\end{center}
\caption{(Color Online) Left panel: Event-averaged shear-stress tensor in the local rest frame of the fluid cell located at x=y=2.625 fm, z=0 fm. Results with $b_{\pi }=5$ are in red, $b_{\pi }=10$ in dark green and $b_{\pi }=20$ in light green. Right panel: Dilepton elliptic flow as a function of $\tau_\pi$ at $\sqrt{s_{NN}}=200$ GeV and 20-40\% centrality class.}
\label{fig:pi_munu_tau_pi_v2_dil}
\end{figure}

We can already predict qualitatively the effects of $\tau_\pi$ on the evolution of $\pi^{\mu\nu}$ from: (i) Eq. (\ref{eq:pi_munu_evol}) and (ii) using the fact that the initial conditions of $\pi^{\mu\nu}(\tau_0)=0$. Indeed, since Eq. (\ref{eq:pi_munu_evol}) is a relaxation equation, increasing $\tau_\pi$ (with $\pi^{\mu\nu}(\tau_0)=0$) postpones the development of $\pi^{\mu\nu}$ at early times, thus making the system behave more like ideal fluid with $T^{\mu\nu}\rightarrow T^{\mu\nu}_0$. This behaviour is explicitly confirmed in the left panel of Fig. \ref{fig:pi_munu_tau_pi_v2_dil}. Thus, any viscous effects at early times are supposed to be suppressed with larger $\tau_\pi$, which will affect the development of anisotropic flow harmonics.

We have computed the anisotropic flow harmonics of thermal dileptons according to the scalar product method:
\begin{equation}
v^{\gamma^*}_n=\frac{\left\langle v^{h}_{n} v^{\gamma^*}_{n} \cos \left[n\left(\Psi^{\gamma^*}_{n}-\Psi^h_{n} \right)\right] \right\rangle_{\rm ev}}{\sqrt{\left\langle (v^h_{n})^2 \right\rangle_{\rm ev}}}
\label{eq:vnSP}
\end{equation}
where $\langle \ldots \rangle_{\rm ev}$ is an average over events. The $v^s_n$ and $\Psi^s_n$ in single event are given by
\begin{equation}
v^s_n e^{i n \Psi^s_n} = \frac{\int d p_T  dy d\phi p_T \left[ p^0 \frac{d^3 N^s}{d^3 p} \right] e^{i n\phi}}{\int d p_T dy  d\phi p_T \left[ p^0 \frac{d^3 N^s}{d^3 p} \right]}\label{eq:vn_psin}
\end{equation}
where $p^0 d^3 N^s/d^3 p$ (or $d^4 N/d^4 p$ for a virtual particle) is the single-particle distribution of particle species $s$. We have chosen $s$ to be all charged hadrons up to a mass of 1.3 GeV.  

Looking back at Eq. (\ref{eq:pi_munu_evol}), anisotropic flow $v^{\gamma^*}_n$ develops faster for high $p_T$ dileptons as $\tau_\pi$ increases (see the right panel of Fig. \ref{fig:pi_munu_tau_pi_v2_dil}), since these high energy dileptons mostly originate from the QGP, where the size of the $\pi^{\mu\nu}$ is smaller with increasing $\tau_\pi$. 
The medium having the larger $\tau_\pi$ will, however, become more viscous at later times, i.e. will have a larger $\pi^{\mu\nu}$ (see left panel of Fig. \ref{fig:pi_munu_tau_pi_v2_dil}). Thus low $p_T$ thermal dileptons will have smaller flow as $\tau_\pi$ increases (see right panel of Fig. \ref{fig:pi_munu_tau_pi_v2_dil}). Therefore, it may be possible to constrain $\tau_\pi$ from dilepton flow data provided the statistical/systematic uncertainties are sufficiently small. Note that hadronic $v_2$ is essentially insensitive to $\tau_\pi$ \cite{Vujanovic:2014vwa}. 

\begin{figure}[!h]
\begin{center}
\includegraphics[width=0.496025\textwidth]{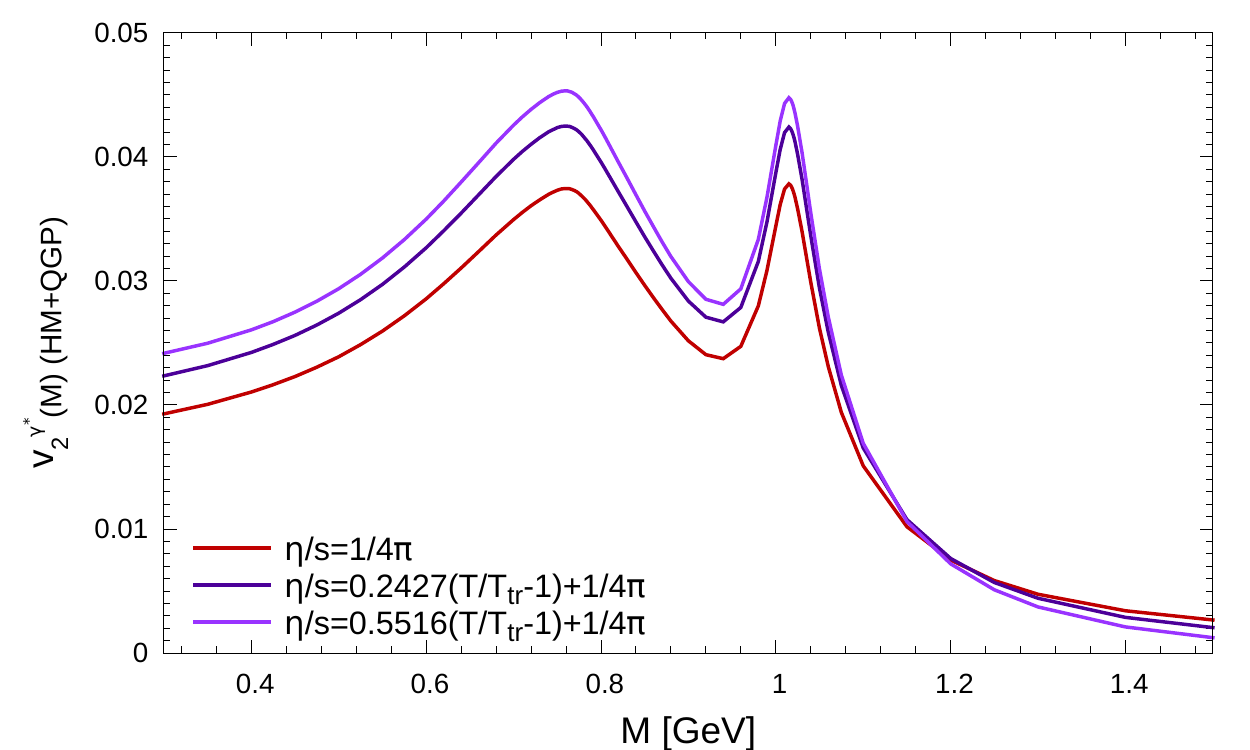}
\includegraphics[width=0.496025\textwidth]{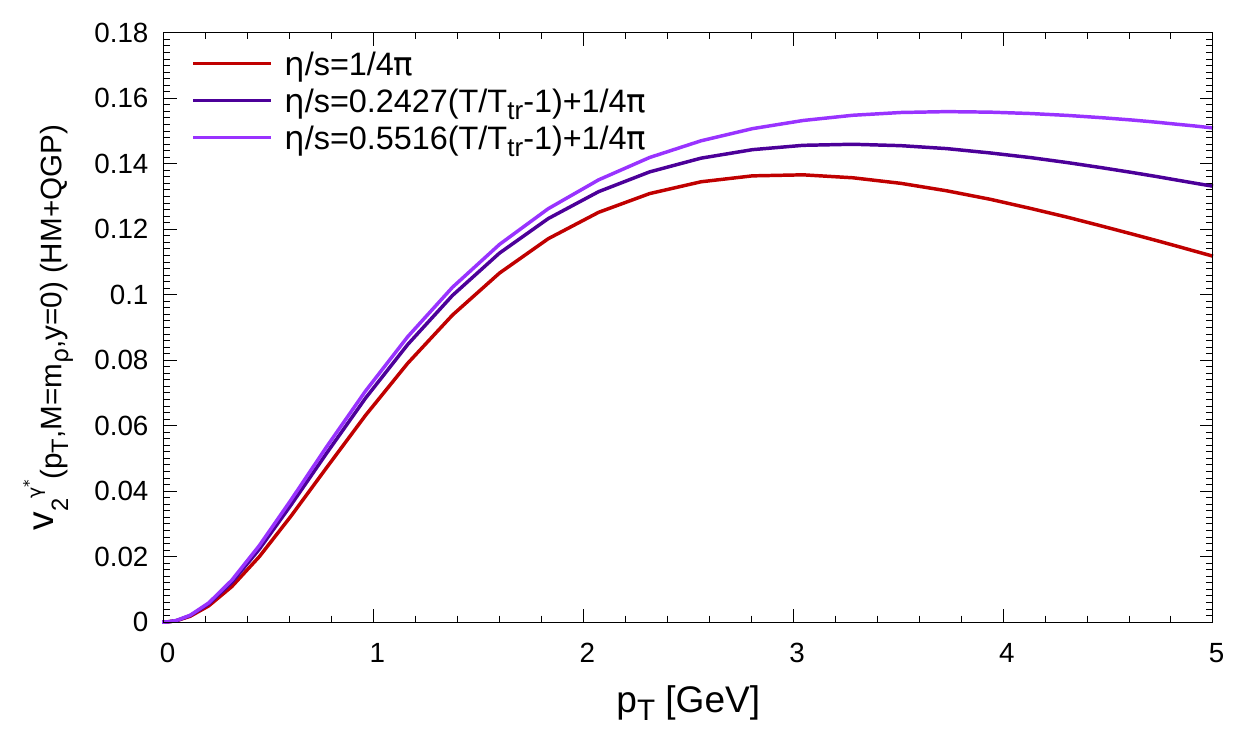}
\includegraphics[width=0.496025\textwidth]{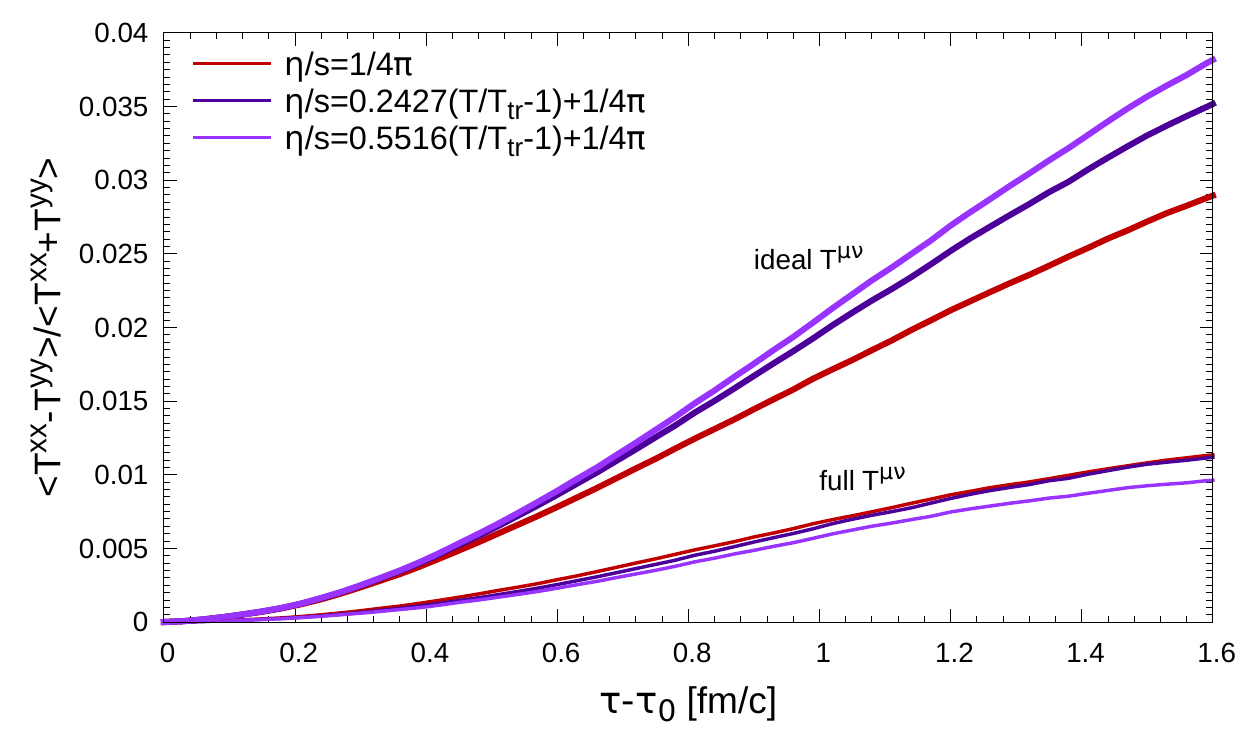}
\includegraphics[width=0.496025\textwidth]{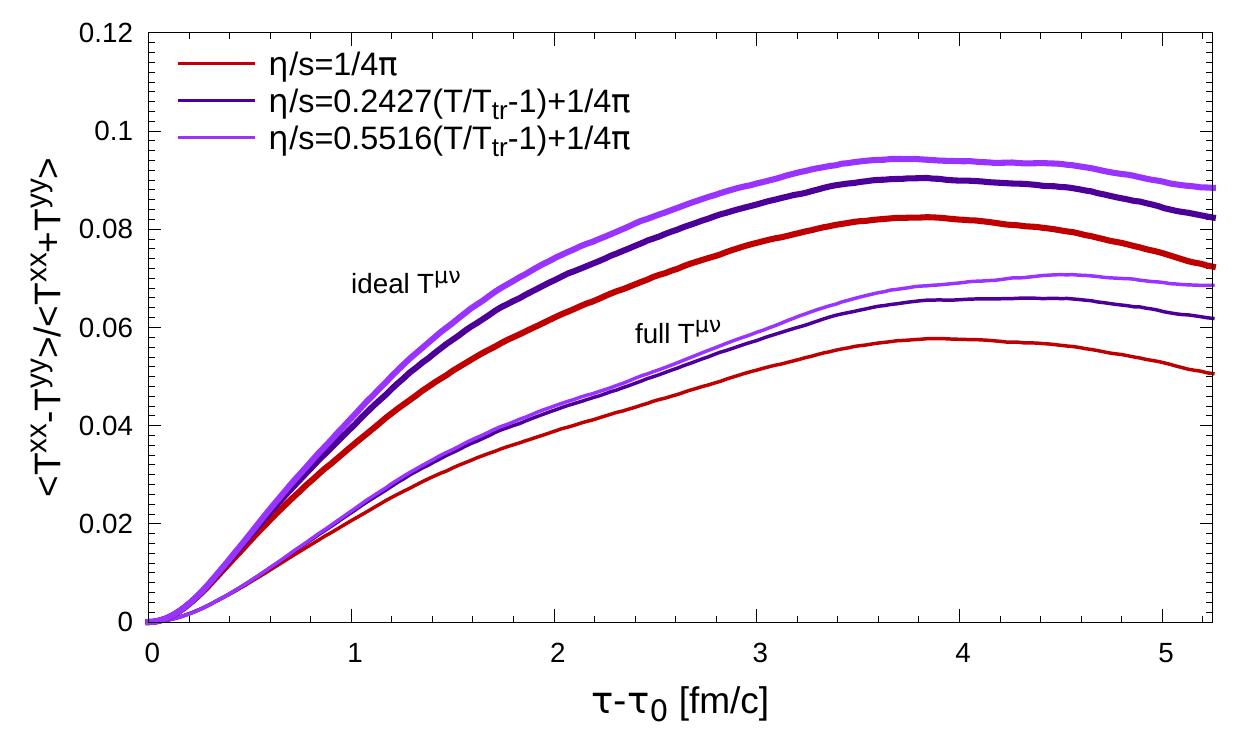}
\end{center}
\caption{(Color Online) $v_2$ of thermal dileptons as a function of invariant mass (top left panel) and as function of $p_T$ (top right panel) at $\sqrt{s_{NN}}=200$ GeV and 20-40\% centrality class. Momentum anisotropy $\epsilon_p=\langle T^{xx}-T^{yy}\rangle/\langle T^{xx} + T^{yy}\rangle$ as a function of time for the QGP phase alone (bottom left panel) and the HM phase alone (bottom right panel). $\epsilon_p$ was computed using solely $T^{\mu\nu}_0$ (see label ideal $T^{\mu\nu}$) and the full $T^{\mu\nu}$ (see label).}
\label{fig:v2_eta_s_T}
\end{figure}

Thermal dileptons are also quite sensitive to the temperature dependence of $\frac{\eta}{s}$ as can be seen in Fig. \ref{fig:v2_eta_s_T}, where the $v_2$ of thermal dileptons was again computed using the scalar product method. 
Note that the ordering of the curves follows closely the momentum anisotropy $\epsilon_p\equiv \langle T^{xx}-T^{yy}\rangle/\langle T^{xx}+T^{yy}\rangle$ of the system. The physical reason behind the ordering of the $\epsilon_p$ curves in Fig. \ref{fig:v2_eta_s_T} will be explored in an upcoming publication. The goal here is to show that the $v_2$ of thermal dileptons is sensitive to the slope of the linear temperature dependence of $\frac{\eta}{s}$ chosen, while the $v_2$ of charged hadrons isn't, see \cite{Vujanovic:2014vwa}. This statement of course only holds true for the top collision energy at RHIC. Hence thermal dileptons can be used to study the temperature dependence of $\frac{\eta}{s}$ in the QGP phase, and even in the transition between QGP and HM, at RHIC. 

\section{Dileptons from the Beam Energy Scan}
\begin{figure}[!h]
\begin{center}
\includegraphics[width=0.496025\textwidth]{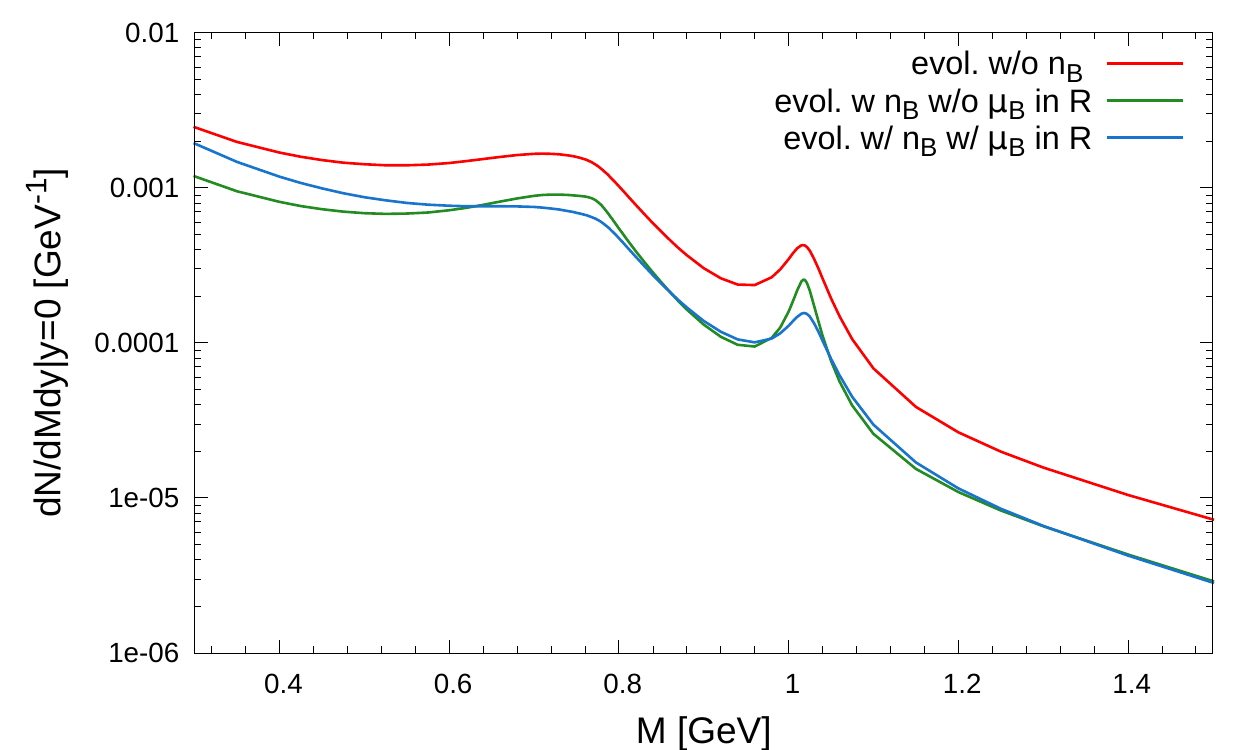}
\includegraphics[width=0.496025\textwidth]{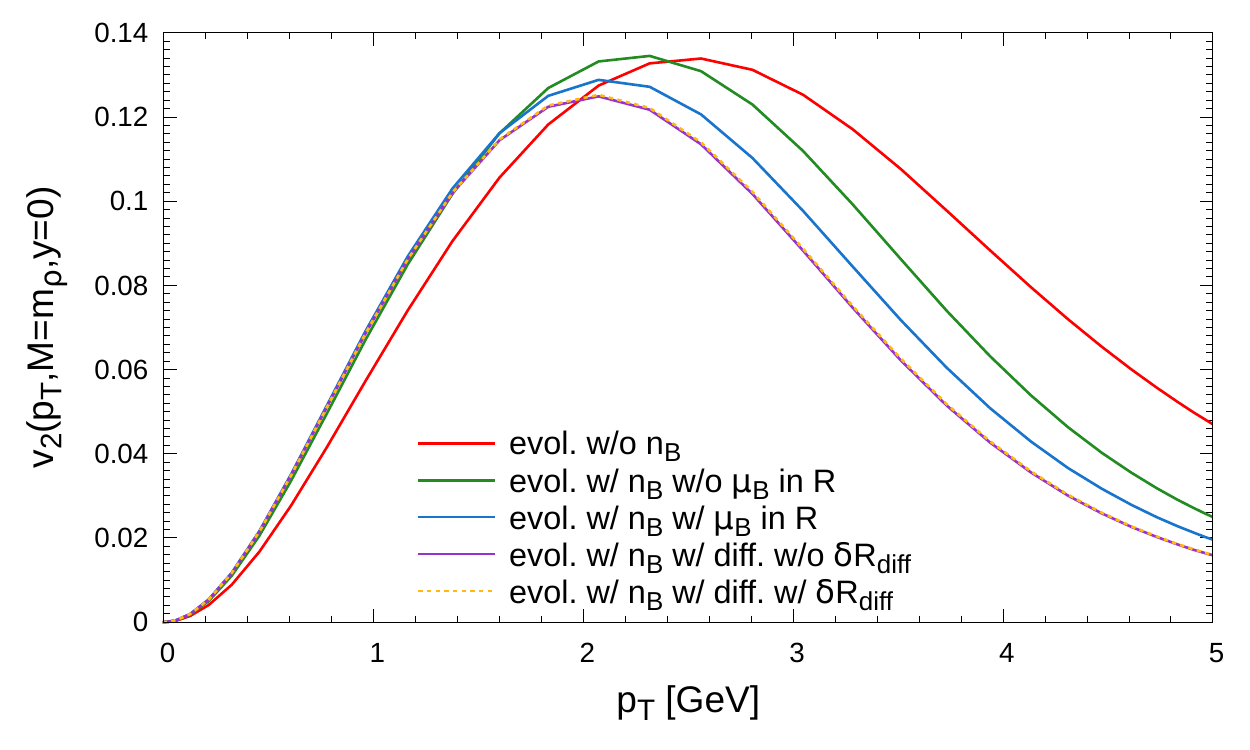}
\end{center}
\caption{(Color Online) Dilepton yield (left panel) and $v_2$ (right panel) at $\sqrt{s_{NN}}=7.7$ GeV and for 0-80\% centality class.}
\label{fig:BES}
\end{figure}

The dilepton results from the BES will be presented as follows: effects of an equation of state with finite $\mu_B$ will be studied first before investigating the influence of baryon diffusion, and thus baryon conductivity $\kappa$, on dilepton production. The dilepton yield and $v_2$ are presented in Fig. \ref{fig:BES}. The red curve simulates a medium at collision energy $\sqrt{s_{NN}}=7.7$ GeV using solely a baryon-free EoS. As baryon chemical potential is turned on in the EoS, while leaving $\mu_B=0$ in dilepton rates, the green curve is obtained. The sharp drop in dilepton yield between the red and greed curve is caused by the significant change in the temperature profile owing to the presence of the $\mu_B$ degree of freedom. Inserting $\mu_B$ in dilepton rates yields (blue curve) causes a significant in-medium broadening of vector mesons as is clearly seen in the dilepton yield. 
Baryon diffusion doesn't generate much entropy production nor does in change $\mu_B$ or $T$, hence the dilepton yield with baryon diffusion isn't significantly changed and is thus not plotted in the left panel of Fig. \ref{fig:BES}. However, there is a substantial change in the $v_2$ of thermal dileptons owing to baryon diffusion (see purple/yellow curves in the right panel of Fig. \ref{fig:BES}).  This result is promising as it opens the possibility to constrain $\kappa$ using thermal dileptons. 
\section{Conclusion}
In summary, we have shown tha thermal dileptons are sensitive observables to various transport coefficients of dissipative hydrodynamics. Thus, one should not solely rely on hadronic production to constrain the various transport coefficients of dissipative hydrodynamics: a combination of hadronic and electromagnetic probes (of which dileptons are a particular category) will yield much improved constraints; a more robust extraction of transport coefficients thus becomes possible. 

\Acknowledgements G. Vujanovic acknowledges support by the Canadian Institute for Nuclear Physics and the Schulisch Graduate Fellowship from McGill University. G.S. Denicol acknowledges support through a Banting Fellowship from the Government of Canada. Computations were performed on the Guillimin supercomputer at McGill University under the auspices of Calcul Qu\'ebec and Compute Canada. The operation of Guillimin is funded by the Canada Foundation for Innovation (CFI), the National Science and Engineering Research Council (NSERC), NanoQu\'ebec, and the Fonds de Recherche du Qu\'ebec | Nature et Technologies (FRQNT).

\bibliographystyle{elsarticle-num}
\bibliography{references}

\end{document}